\documentclass[conference]{IEEEtran}
\usepackage{amsmath,amsfonts}
\usepackage{amssymb}
\usepackage{mathrsfs}   % For \mathscr
\usepackage{algorithmic}
\usepackage{algorithm}
\usepackage{array}
\usepackage[
    caption=false,
    font=footnotesize,
    labelfont=sf,
    textfont=sf
]{subfig}
\usepackage{textcomp}
\usepackage{stfloats}
\usepackage{url}
\usepackage{verbatim}
\usepackage{graphicx}
\usepackage{cite}
\def\BibTeX{{\rm B\kern-.05em{\sc i\kern-.025em b}\kern-.08em
    T\kern-.1667em\lower.7ex\hbox{E}\kern-.125emX}}
\usepackage{balance}
\usepackage{bm}         % For \bm
\usepackage{cite}
\usepackage{booktabs}
\usepackage{multirow}
\usepackage{tabularx} % Allows auto-resizing columns
\usepackage{xcolor}
\usepackage{chngcntr}
\counterwithout{figure}{section}
\usepackage{threeparttable}

\begin{document}
\title{Grid Demand Flexibility Assessment of AI Data Centers via Batch Workload Temporal Shifting}
% \author{Suntao~Su,~Liang~Du,~and~Junbo~Zhao
% \thanks{S. Su and L. Du are with the Department of Electrical and Computer Engineering, Villanova University, Villanova, PA 19085, USA (email: ssu02@villanova.edu; liang.du@villanova.edu).}
% \thanks{J. Zhao is with the Department of Electrical and Computer Engineering, University of Connecticut, Storrs, CT 06269, USA. J. Zhao is also with Dartmouth College, Hanover, NH 03755, USA (email: junbo@uconn.edu).}}
\author{\IEEEauthorblockN{Suntao Su, Liang Du}
\IEEEauthorblockA{\textit{Dept. of Electrical \& Computer Engineering} \\
\textit{Villanova University}\\
Villanova, PA USA \\
ssu02@villanova.edu, liang.du@villanova.edu}
\and
\IEEEauthorblockN{Shengyi Wang}
\IEEEauthorblockA{\textit{Dept. of Electrical \& Computer Engineering} \\
\textit{University of Arkansas at Little Rock}\\
Little Rock, AR USA \\
swang@ualr.edu}
}
\maketitle

\begin{abstract}
The rapid growth of artificial intelligence (AI) data centers has introduced new challenges to power system operation. As their power demand becomes larger and more variable, quantitatively characterizing their demand flexibility is increasingly important for effective power system coordination. However, heterogeneous workload characteristics and resource requirements make this flexibility difficult to characterize directly. This paper proposes a framework for assessing the grid-compatible demand flexibility of AI data centers via batch workload temporal shifting. An averaging-based resource usage processing method is developed to map fine-resolution CPU, GPU and memory usage into unified time intervals compatible with power system operation. A workload temporal scheduling model is then formulated to shift batch workloads while preserving execution continuity, delay constraints, and server resource capacities, and is coupled with a utilization-dependent server power model to translate workload scheduling decisions into server power demand. Two complementary flexibility metrics are evaluated: short-term peak demand shaving and the maximum duration of sustained power reduction. Numerical results based on real GPU cluster traces demonstrate that workload temporal shifting can provide quantifiable and grid-compatible demand flexibility for AI data centers with limited disruption to computing workloads.
\end{abstract}

\IEEEpeerreviewmaketitle

\section{Introduction}

The rapid development of artificial intelligence (AI) has led to a substantial increase in the deployment of large-scale data centers equipped with high-performance GPUs. As computing demand continues to grow, the power consumption of data centers is becoming increasingly significant and time-varying. Modern hyperscale AI data centers can require power capacities exceeding 100 MW, while some future facilities are expected to reach the gigawatt scale. Highly dynamic AI workloads can introduce rapid and substantial power fluctuations that pose new challenges to power system operation \cite{chen2025electricity}. 
% This trend further strengthens the coupling between computing systems and power systems \cite{wu2026digital}. In such a coordination paradigm, data centers are no longer regarded only as passive electricity consumers, but also as controllable demand-side resources. 

At the same time, data centers also offer new opportunities for power system operation as flexible demand-side resources. Emerald AI, Google and NVIDIA launched the AI Energy Management Alliance \cite{parker2026aema}, which promotes flexible AI data centers that adjust electricity demand through workload shifting, energy storage, and on-site generation to support grid reliability and accelerate interconnection \cite{biggo2026aienergy}. Experimental studies have also demonstrated that software-based workload orchestration can provide rapid and sustained grid-responsive flexibility in GPU-based AI data centers, including a 25\% power reduction for three hours while maintaining workload quality of service \cite{williams2026power,colangelo2026ai}.

Existing data center flexibility strategies can generally be divided into two main categories. On the one hand, several studies have demonstrated the inherent flexibility of supporting infrastructures. A genetic-algorithm-based optimization method for cold-plate liquid-cooled data centers is proposed in \cite{qu2024real}, which jointly adjusts cooling-tower airflow and coolant flow rates under chip-temperature constraints. Reference \cite{mohammadi2026grid} reviews AI data-center load characteristics, highlighting the role of coordinated server-, rack-, and grid-level energy storage in smoothing demand and supporting grid integration.

On the other hand, extensive research has focused on information technology (IT)-based strategies. In \cite{long2025flexible}, IT server power consumption is modeled as the sum of static power and dynamic power proportional to the cube of operating frequency and server utilization, enabling load flexibility through frequency adjustment under service-delay constraints. References \cite{han2025evaluating} and \cite{zhou2026profit} propose frameworks for evaluating the dispatchable capacity of cross-regional data centers by explicitly modeling spatial flexibility through interregional workload transfers while accounting for spatiotemporal scheduling coupling and data transmission constraints. Reference \cite{caprara2026data} estimates IT-side GPU power using a linear model, and proposes a latency-aware task-deferral strategy to provide demand response flexibility.  

However, two limitations remain: (1) existing flexibility assessment studies do not sufficiently address the temporal resolution mismatch between fine-grained computing workload traces and the coarser intervals used in power system operation, limiting their direct integration into grid scheduling and control. (2) existing studies either focus on servers dominated by CPU or GPU resource \cite{caprara2026data},\cite{wang2025demand}. Consequently, the combined effects of CPU, GPU, and memory resource utilization on workload flexibility have not been fully characterized.

To address these limitations, this paper proposes a comprehensive workload flexibility assessment framework that bridges heterogeneous computing workloads and grid-side demand flexibility. An averaging-based processing method is proposed to transform fine-resolution resource traces into modular data compatible with power system scheduling and operation time intervals. A workload power flexibility assessment model is formulated by integrating batch workload scheduling, multi-server resource capacity constraints, and a utilization-dependent server power model. The model directly maps workload scheduling decisions into power consumption.

\section{Modular Data of Workloads}

This section first introduces the potential temporal flexibility in workloads, then details the averaging-based resource usage processing method.

\subsection{Potential Temporal Flexibility in Workloads}

Due to heterogeneous workload characteristic and service requirements, not all jobs are suitable for temporal shifting. Workloads can generally be divided into two categories: online and offline jobs. Specifically, online jobs are typically delay-sensitive and have limited allowable shift time intervals, including interactive web services, video streaming, and online
gaming. In contrast, offline jobs can tolerate longer waiting or processing times, typical examples include data backup, offline AI model training, and scientific simulations, etc. Such workloads are referred to as batch workloads, and only their flexibility is quantified in this paper.

\begin{figure}[t]
\centering
\includegraphics[width=3.5in]{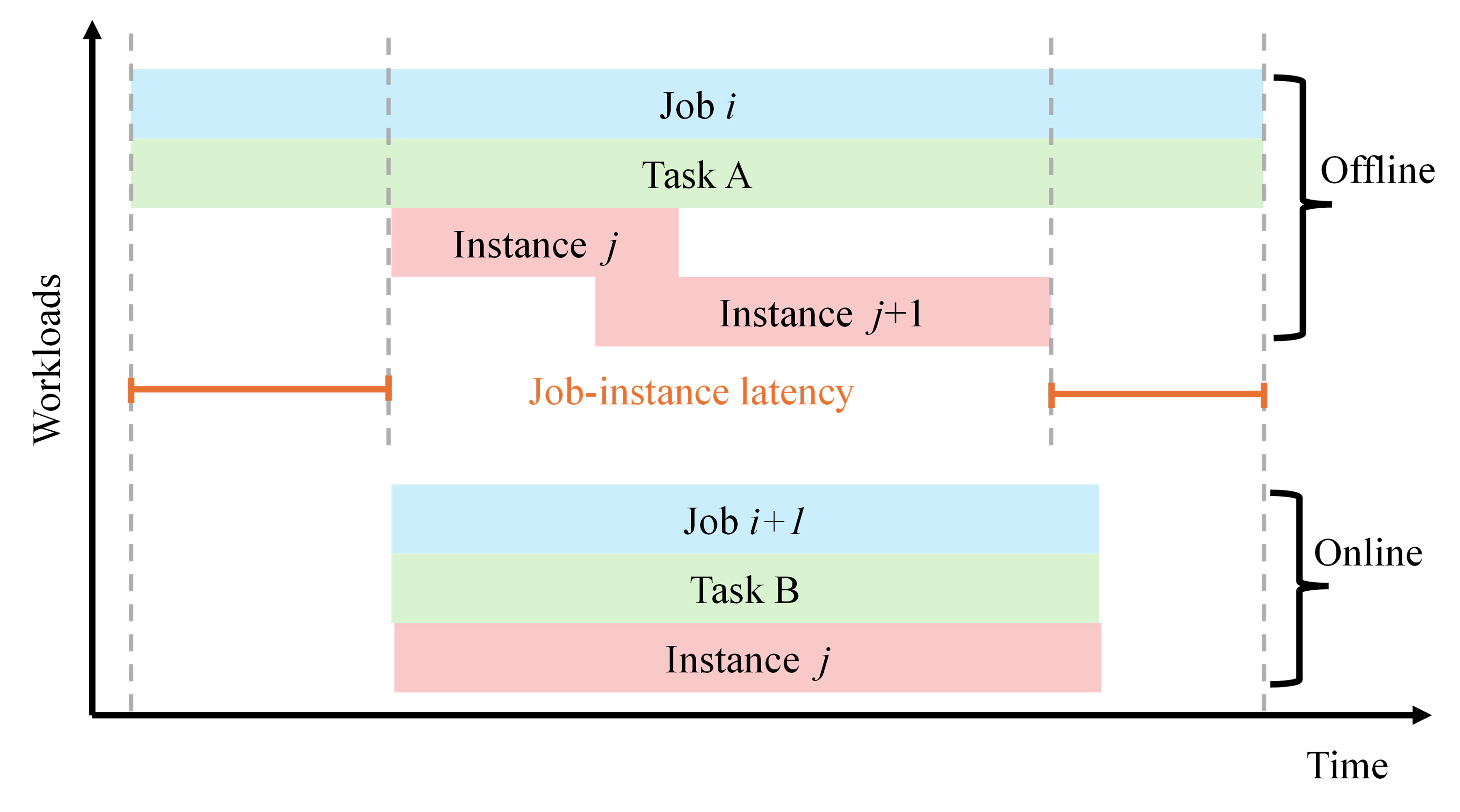}
\caption{{
\raggedright
The illustration of the hierarchical temporal relationship among jobs, tasks, and instances for online and offline jobs.
}}
\label{fig1}
\end{figure}

Let $\mathcal{I}$ and $\mathcal{J}_i$ denote the set of jobs and instances in job $i$, respectively. The hierarchical temporal relationship among jobs, tasks, and instances is shown in Fig. \ref{fig1}. As shown in this figure, after an offline job $i$ is submitted, the instances may not be executed immediately, resulting in a waiting interval between job submission and instance execution. The orange interval represents the job-instance latency, which provides the allowable temporal flexibility. Thus, from the operators' perspective, the execution interval of a batch workload can be shifted as long as the rescheduled execution remains within its allowable time window. This allows the required jobs to be completed within their predefined time limits while providing temporal flexibility to the power system.

\subsection{Deferrable Jobs}

Grid scheduling and market models commonly operate at time resolutions such as 5, 15, or 60 min, which are not directly aligned with the second-level temporal resolution of the original workload dataset. This mismatch makes the workload traces difficult to directly use for power-system scheduling and demand-response applications. Therefore, the workload data are transformed into modular time intervals consistent with the temporal resolution required for grid operation.
%The maximum allowable shift time interval of workloads can be derived:
% \begin{equation}
%     D_i =
%      \max(\bm{t}_{i}^{\mathrm{ins\_e}})-\min(\bm{t}_{i}^{\mathrm{ins\_s}}) + 1,
% \end{equation}
% \begin{equation}
%     \lambda_i = t_{i}^{\mathrm{job\_e}}-
%      t_{i}^{\mathrm{job\_s}} - D_i + 1,
% \end{equation}
% where $D_i$ represents the actual execution time interval of job $i$, which can be computed based on the start and end time intervals of instances; $\bm{t}_{i}^{\mathrm{ins\_s}}$ and $\bm{t}_{i}^{\mathrm{ins\_e}}$ are the starting and end time interval vectors of all corresponding instances in the $i$th job, respectively; $\lambda_i$ is the maximum allowable shift time interval of job $i$; 

We define the time resolution as $\delta$. Accordingly, each day is divided into $T=(24\times60)/\delta$ uniform time intervals $\mathcal{T}=\{1,2,\ldots,T\}$. A job is regarded as a deferrable batch workload if its execution is fully contained within the daily scheduling horizon and its maximum allowable shifting time interval $\lambda_i$ is at least one time interval.
\begin{align}
\text{Deferrable}_i \iff
\left\{
\begin{aligned}
a_i^{\mathrm{job\_s}} \ge 1,
b_i^{\mathrm{job\_e}} \le T,
\lambda_i \ge 1.
\end{aligned}
\right\},
\end{align}
where $a_{i}^{\mathrm{job\_s}}$ and $b_{i}^{\mathrm{job\_e}}$ are the start and end times of the $i$th job, respectively.

The remaining jobs within the day are regarded as online workloads. To limit excessive workload delays and maintain acceptable service quality, we introduce a delay rate $\gamma$, which allows a batch workload to be delayed by up to an additional $\gamma D_i$ intervals beyond its original allowable window, where $\gamma$ can be set based on different service requirements. Thus, we can derive the new maximum allowable shifting time interval
\begin{equation}
    \lambda^\mathrm{new}_i = \lambda_i + \gamma D_i.
\end{equation}

% For each day $d$, the job interval matrix is defined as
% \begin{equation}
%     \bm{S}^{d} \in \{0,1\}^{m\times n},
% \end{equation}
% where each row corresponds to a job and each column corresponds to a $\delta$-min time interval; $m$ is the number of elements of set $\mathcal{I}_d$; $n$ is the number of elements of set $\mathcal{W}$. The element $S_{i,w}^{d}$ is defined by
% \begin{equation}
%     S_{i,w}^{d} =
%     \begin{cases}
%     1, & \text{if job } i \text{ is within time interval } w,\\
%     0, & \text{otherwise}.
%     \end{cases}
% \end{equation}

\subsection{Averaging-based Resource Usage Processing Method}
After identifying the deferrable jobs and defining the unified time resolution \(\delta\), the original resource usage traces must be mapped to the corresponding scheduling intervals. Thus, the averaging-based data processing method is proposed.
% and illustrated in Fig. \ref{fig2} using a  simple example of one job with two instances.

AI workloads typically consume multiple computing resources, including CPU, GPU, and memory. We assume that the resource usage attributed to an instance is constant during its execution time and equal to that allocated by the scheduler. For each instance $j$, the overlap between workload execution interval and each time interval $t$ is calculated as
\begin{equation}
    \Delta _{j,t} =
    \min(b_{j}^{\mathrm{ins\_e}},t\delta)
    -
    \max(a_{j}^{\mathrm{ins\_s}},(t-1)\delta),
\end{equation}
where $a_{j}^{\mathrm{ins\_s}}$ and $b_{j}^{\mathrm{ins\_s}}$ are the start and end times of the instance $j$, respectively.

The averaged resource usage of instance $j$ at time interval $t$ is then computed as
\begin{equation}
    n_{j,t}^{{re}} = n_{j}^{{re}}\Delta_{j,t}/{\delta},
\end{equation}
where $re$ denotes the one of the resource type of set $\mathcal{RE}=\{\mathrm{CPU}, \mathrm{GPU}, \mathrm{memory}\}$.

Lastly, if multiple instances of the same job overlap with the same time interval, they would be aggregated. Therefore, the averaged resource usage of job $i$ at time interval $t$ can be derived as follows
\begin{equation}
    n_{i,t}^{{re}} = \sum \nolimits_{j\in \mathcal{J}_i} {n}_{j,t}^{{re}}.
\end{equation}

% \begin{figure}[t]
% \centering
% \includegraphics[width=3.5in]{Figure/aggregated instance-level usage.png}
% \caption{\parbox[t]{\linewidth}{
% \raggedright
% The averaging-based data processing for workload usage.
% }}
% \label{fig2}
% \end{figure}

\section{Flexibility Model of Batch Workloads}
In this section, the batch workload flexibility model is proposed, including the batch workloads shifting model and the data center IT power consumption model.

\subsection{Batch Workloads Shifting Model}
A batch workload $i$ is described by $\{t_{i,s},t_{i,e},D_{i}, \lambda_i^\mathrm{new}, \bm{n}_{i}^\mathrm{CPU},\bm{n}_{i}^\mathrm{GPU},\bm{n}_{i}^\mathrm{memory}\}$, denoting its submission time interval, end time interval, execution duration, maximum allowable shifting time interval, and CPU, GPU and memory resource demand, respectively. For workload $i$, its execution schedule is represented by a binary variable $z_{i,k}$, where $z_{i,k}=1$ indicates that job $i$ starts at feasible time interval $k$, and $z_{i,k}=0$ otherwise.
 
Considering the non-preemptive nature of the workloads, once a workload starts execution, it cannot be interrupted. Thus, each job must select exactly one feasible starting time, $z_{i,k}$ should satisfy the following constraint:
\begin{equation}
    \sum \nolimits_{k\in \mathcal{K}_i}z_{i,k}=1,~~~ \forall i \in \mathcal{I}_\mathrm{off},
\end{equation}
where $\mathcal{K}_i=\{0,1,...,\lambda_i^\mathrm{new}\}$ is the set of all feasible starting times of job $i$; $\mathcal{I}_\mathrm{off}$ represents the set of all batch workloads.

% The CPU and GPU resource demands after shifting are expressed as:
% \begin{align}
% \widetilde{\bm{n}}_i^\mathrm{CPU} &= \sum_{j=0}^{\lambda_i}z_{i,j} S_j \bm{n}_i^\mathrm{CPU},\\
% \widetilde{\bm{n}}_i^\mathrm{GPU} &= \sum_{j=0}^{\lambda_i}z_{i,j} S_j \bm{n}_i^\mathrm{GPU},
% \end{align}
% where $\bm{n}_i^\mathrm{CPU}$ and $\bm{n}_i^\mathrm{GPU}$ are the CPU and GPU resource demand vectors of the workload $i$ during the job duration time, respectively; and $\sim$ represents the status after shifting; $S_i$ is the matrix of coefficients for the shifting process. For example, the CPU resource demand vector of a workload is $[2,1,0,0]$, and we expect it starts in second time slot, i.e., $[0,2,1,0]$. Thus, we have
% \begin{equation}
% \begin{bmatrix}
% 0\\
% 2\\
% 1
% \end{bmatrix}
% =
% \left(0
% \begin{bmatrix}
% 1 & 0 & 0\\
% 0 & 1 & 0\\
% 0 & 0 & 1\\
% \end{bmatrix}+
% 1
% \begin{bmatrix}
% 0 & 0 & 0\\
% 1 & 0 & 0\\
% 0 & 1 & 0\\
% \end{bmatrix}+
% 0
% \begin{bmatrix}
% 0 & 0 & 0\\
% 0 & 0 & 0\\
% 1 & 0 & 0\\
% \end{bmatrix}\right)
% \begin{bmatrix}
% 2\\
% 1\\
% 0
% \end{bmatrix}
% \end{equation}

\subsection{Power Consumption Model of Servers}
To evaluate the potential flexibility of servers, an estimated power consumption is required. After job submission, the schedulers dynamically assigns workloads to servers with sufficient available resources. In this paper, we assume that job schedulers use the predefined scheduling algorithm. The resource demand assigned to server $s$ at time interval $t$ consist of two components: online and offline workloads.
\begin{equation}
    u^{re}_{s,t=k+r} = \sum \nolimits_{i \in \mathcal{I}_\mathrm{off}}{w^{re}_{s,i,r}n_{i,r}^{re}z_{i,k}} + 
    \sum \nolimits_{i \in \mathcal{I}_\mathrm{on}} u_{s,i,r}^{re},
\end{equation}
where $w^{re}_{s,i,r}$ denotes weights of job $i$'s resource demand at time interval $r$ assigned to server $s$; $r\in \mathcal{R}=\{1,...,D_{i}\}$ denotes the relative execution interval of job $i$.

The CPU, GPU and memory usage of each server must not exceed its corresponding resource capacity. The resource constraints in job scheduling decisions are described below. 
\begin{align}
    0 \leq u_{s,t}^{{re}}
    &\leq
    N^{{re}}_s,
\end{align}
where $N^{{re}}_s$ is the available resource capacity of server $s$.

For conventional CPU-dominated data centers, server power consumption is often estimated primarily from CPU utilization \cite{ma2024data}. However, GPU utilization cannot be omitted in AI data centers, thus, we adopt the linear server power model in which power is approximately proportional to the utilization of CPU/GPU/memory resources. The objective is to obtain a workload-level estimate of electrical demand rather than a hardware-accurate server power model. Thus, the instantaneous power consumption of server at time interval $t$ can be calculated as follows
\begin{equation}
    {P}_{s,t}^\mathrm{ser} = {P}_{s}^\mathrm{idle} +
    \sum \nolimits_{re \in \mathcal{RE}}{C}_{s}^{re}  {u_{s,t}^{re}}/{N^{{re}}_s},
\end{equation}
%{P}_{s}^\mathrm{online} + ${P}_{s}^\mathrm{online}$ is the power consumption of the online workload, which can be foretasted using the historical data;
where ${P}_{s,t}^\mathrm{ser}$ is the power consumption of server $s$; ${P}_{s}^\mathrm{idle}$ is the idle power of server $s$; $C_s^{re}$ is the rated resource power.

Finally, aggregating all physical servers to obtain the data center IT power load at time interval $t$ as follows
\begin{equation}
    P^{\mathrm{IT}}_{t}
    =
    \sum \nolimits_{s \in \mathcal{S}_t}
    P^{\mathrm{ser}}_{s,t},
\end{equation}
where $\mathcal{S}_t$ is the set of active servers at time interval $t$.

% \subsubsection{Model of Auxiliary Storage Device}
% Grid-interactive UPS (GiUPS) systems can respond quickly to
% disturbances and assist with frequency regulation or voltage
% ride through in AI data center. Unlike traditional UPS systems, which operate independently of grid conditions
% except during outages, GiUPS systems continuously monitor
% grid conditions and adjust their output accordingly. The energy storage constraints of the installed GiUPS are formulated as follows:
% \begin{equation}
% E_{t}^{\mathrm{UPS}}
% = E_{t-1}^{\mathrm{UPS}}
% + \eta_{\mathrm{ch}} P_{t}^{\mathrm{ch}}
% - \frac{P_{t}^{\mathrm{dis}}}{\eta_{\mathrm{dis}}}
% \end{equation}
% \begin{equation}
% 0\le P^\mathrm{ch}_{d,t}\le P^\mathrm{ch,\max}_d
% \end{equation}
% \begin{equation}
% 0\le P^\mathrm{dis}_{d,t}\le P^\mathrm{dis,\max}_d
% \end{equation}
% \begin{equation}
% 0\le E_{d,t}\le E^\mathrm{\max}_d
% \label{eq2:dc model}
% \end{equation}
% where $E_{t}^{\mathrm{UPS}}$ is the energy stored in electricity from the GiUPS at time $t$; $P_{t}^{\mathrm{ch}}$ and $P_{t}^{\mathrm{dch}}$ are the charging and discharging power at time $t$, respectively; $\eta_{\mathrm{ch}}$ and $\eta_{\mathrm{dis}}$ are the corresponding charging and discharging efficiency, respectively; $E_{\mathrm{max}}^{\mathrm{UPS}}$ and ${P}_{\mathrm{max}}^{\mathrm{UPS}}$ are the capacity of stored energy and power of charging or discharging, respectively. 

\section{Grid Demand Flexibility Assessment Framework}

To comprehensively characterize the workload demand flexibility, two optimization frameworks are developed to evaluate two operating capabilities: short-term peak demand shaving and sustained power reduction. 

\subsection{Short-Term Peak Demand Shaving}
Short-term peak demand shaving focuses on the data center’s ability to rapidly reduce its power consumption during a critical high-demand period. The objective is to quantify the maximum grid demand flexibility by scheduling the minimum number of batch workloads. 

Let the requested grid-service event begin at $t_0$ and last for $D^{\mathrm{evt}}$ intervals. Define $\mathcal{T}^{\mathrm{evt}} = \{t_0, t_0+1,..., t_0 + D^{\mathrm{evt}}\}$. The number of controlled jobs in the assessment framework should be as small as possible, thus, we define a binary variable $c_i$, where $c_i=1$ indicates that job $i$ is shifted from its original starting time $k_i^0$, and $c_i=0$ otherwise.

Thus, the optimization problem can be modeled as follows
\begin{align}
    &\min~~~ -  F + \alpha \sum \nolimits_{i \in \mathcal{I}_\mathrm{off}}{c_i} \\
    \mathrm{s.t.}~~~ &(1)-(10),\\
    &c_i = 1 - z_{i,k_i^0},~ \forall i \in \mathcal{I}_\mathrm{off}, \label{cz}\\
    & F \leq P^\mathrm{orig}_t - {P^{DC}_{t}},~ F\ge0,~ \forall t\in \mathcal{T^\mathrm{evt}},
     \label{eq1}
\end{align}
where $F$ is the guaranteed grid-compatible flexibility; $\alpha$ is a coefficient that controls the trade-off, and the value of $\alpha$ is set to 0.01 so that maximizing \(F\) remains the dominant objective, while the number of shifted workloads is minimized as a secondary consideration.; $P^\mathrm{orig}_t$ is the original power load of the workload at time interval $t$, which is decided by the given job arrival traces; constraint (\ref{cz}) depicts the relationship between $c_i$ and $z_i$; constraint (\ref{eq1}) is equivalent to $F = \min \nolimits_{t \in \mathcal{T}^\mathrm{evt}}( P_t^\mathrm{orig} - P_t^\mathrm{DC} )$ at the optimum.

\subsection{Sustained Power Reduction}
The practical value of data-center flexibility depends not only on the magnitude of the power reduction, but also on how long that reduction can be continuously maintained. Therefore, for the sustained duration assessment, the power reduction magnitude is fixed at $F^\mathrm{req}$, while the duration is optimized.

We introduce a binary variable \(y_t\), where \(y_t=1\) indicates that the requested power reduction is provided at time interval \(t\), and \(y_t=0\) otherwise. To ensure that the selected flexibility intervals form a single continuous time block, an additional binary variable \(u_t\) is introduced, where \(u_t=1\) indicates the start of a sustained flexibility period, and \(u_t=0\) otherwise.

Thus, the optimization problem can be modeled as 
\begin{align}
    &\min~~~ -  \sum \nolimits_{t \in \mathcal{T}} y_t + \alpha \sum \nolimits_{i \in \mathcal{I}_\mathrm{off}}{c_i} \\
    \mathrm{s.t.}~~~ &(1)-(10),\\
    &c_i = 1 - z_{i,k_i^0},~ \forall i \in \mathcal{I}_\mathrm{off}, \label{c}\\
    &P_t^{\mathrm{dc}}\le
    P_t^{\mathrm{ori}}-F^{\mathrm{req}}+M(1-y_t),~
    \forall t \in \mathcal{T}, \label{big M}\\
    &y_1 \le u_1,~y_t-y_{t-1} \le u_t,~t=2,...,T, \label{begin}\\
    &\sum \nolimits_{t \in \mathcal{T}} u_t \le 1, \label{u}
\end{align}
where constraint (\ref{big M}) depicts the required power reduction represented using the linear big-M constraint; $M$ is the sufficiently large constant; constraint (\ref{begin}) identifies the beginning of a flexibility period; constraint (\ref{u}) ensures only one continuous flexibility period is allowed over the scheduling horizon.

% Since the SLA requirement can be described as the constrained probability of delaying of job execution, the problem can be modeled as the following chance constraints:
% \begin{equation}
%     \mathrm{Pr}\{VR^\mathrm{delay} \geq \mu \} \leq \epsilon
% \end{equation}
% where $\mu$ is the threshold of $VR^\mathrm{delay}$; $\epsilon$ is the risk level of the chance-constrained optimization, set by SLA requirement. The chance constraint is critical for maintaining acceptable SLA.

% However, we would like to use less delay to achieve the maximum flexibility. Thus, We define $v^\mathrm{delay}_i$ to indicate whether the workload $i$ exceeds its end time:
% \begin{equation}
%     v^\mathrm{delay}_i = 
%     \begin{cases}
%         1, & w_{i,s} + D_i + \sum_{j=0}^{\lambda_i^{new}} z_{i,j}J_i > w_{i,e},\\
%         0, & w_{i,s} + D_i + \sum_{j=0}^{\lambda_i^{new}} z_{i,j}J_i \leq w_{i,e}.
%     \end{cases}
% \end{equation}
% where $J_i=[0,1,...,\lambda_i^\mathrm{new}]$; $v^\mathrm{delay}_i=1$ indicates a delay, and $v^\mathrm{delay}_i=0$ indicates that the workload $i$ meets its end time.

% Then, for a coming workload trace, we can derive the violation rate of delay as:
% \begin{equation}
%     VR^\mathrm{delay} = \frac{\sum_{i=1}^{N} v_{i}^\mathrm{delay}}{N} \leq \mu
% \end{equation}
% where $N$ is the number of workloads; $\mu$ is the threshold of $VR^\mathrm{delay}$.

% \subsection{Lower Level Model}
% The lower level model aims to minimize the number of delays given the power cap determined by the upper level model.
% \begin{equation}
%     \min {VR^\mathrm{delay}}
% \end{equation}

\section{Numerical Results}
In this section, we conduct numerical experiments to validate the effectiveness of the proposed model. The workload traces are obtained from one day of the Alibaba Cluster Trace GPU v2020 dataset \cite{weng2022mlaas}, containing a total of 7,828 jobs. The simulations are implemented in Python using Jupyter 6.3.0 with Gurobi 13.0.3.

\subsection{Temporal and Power Characteristics of AI Workloads}
\begin{figure}[!t]
    \centering
    \includegraphics[width=3.5in]{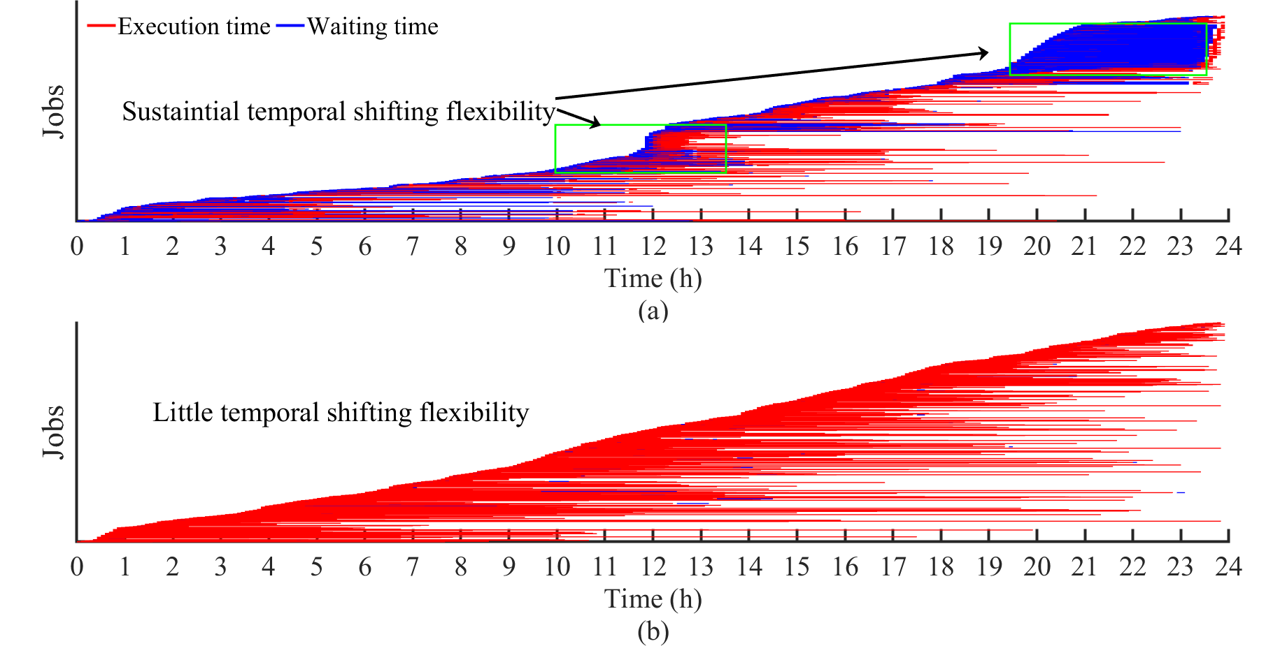}
    \caption{{
    Submission and execution duration time interval of different workload types: (a) batch workload, (b) online workload.
    }}
    \label{fig3}
\end{figure}

The temporal characteristics of batch and online workloads are shown in Fig. \ref{fig3}. As shown in this figure, the blue segments represent the waiting time, while the red segments indicate the execution time. For batch workloads, a considerable number of workloads exhibit non-negligible waiting intervals, indicating substantial temporal scheduling flexibility. In contrast, online workloads are executed almost immediately after submission and therefore contain little or no waiting time. This distinction highlights that batch workloads provide significantly greater temporal flexibility for workload shifting. 

\begin{figure}[!t]
    \centering
    \includegraphics[width=3.5in]{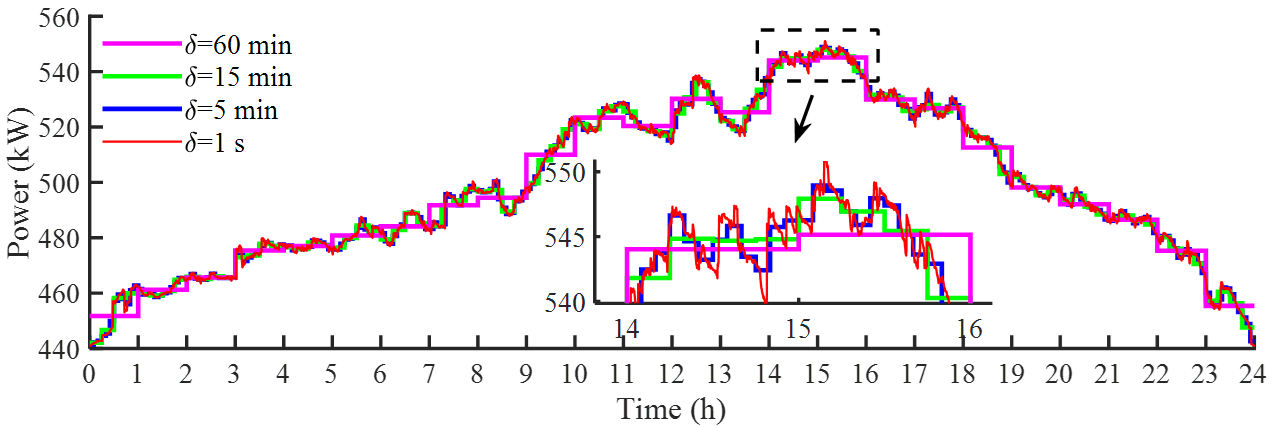}
    \caption{{
    \raggedright
    Daily IT power profiles under different temporal resolutions.
    }}
    \label{fig_total_power}
\end{figure}

In addition, Fig. \ref{fig_total_power} compares the averaged workload power profiles at different temporal resolutions. The fine resolution 1-s profile captures fluctuations in the original computing traces. As $\delta$ increases, the power profiles become smoother, while the main daily load trend and peak characteristics remain well preserved. These results validate that the proposed averaging-based resource usage processing method reduces data granularity while retaining the key power demand characteristics required for power system analysis and scheduling.

\subsection{Peak Power Demand Shaving}
% Peak power demand shaving is selected to evaluate the grid-support capability of data center workload flexibility. By shifting deferrable workloads away from high-demand periods, the data center can reduce short-term peak demand and alleviate stress on the power system without interrupting workload execution.

\begin{figure}[!t]
    \centering
    \includegraphics[width=3.5in]{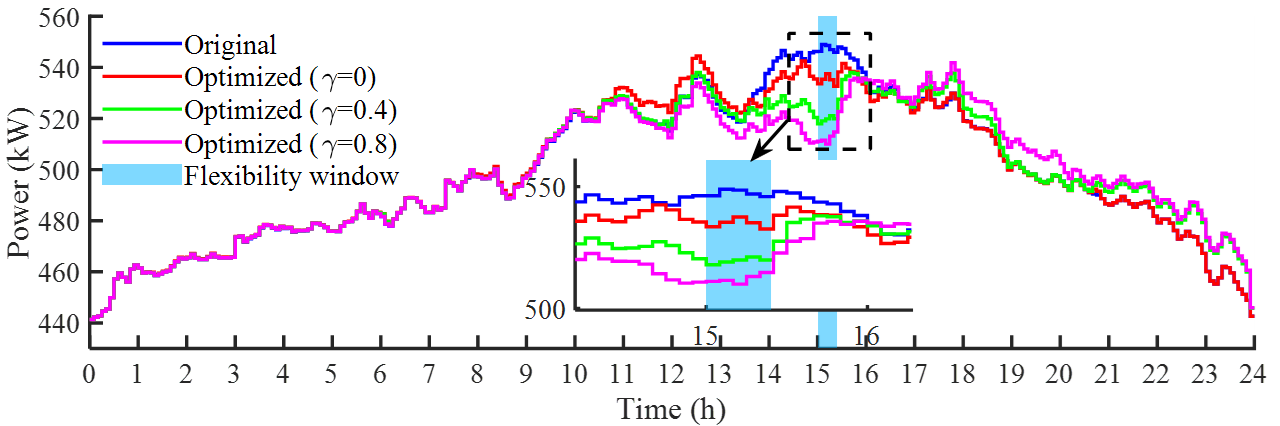}
    \caption{{
    \raggedright
    Daily IT power profiles before and after temporal workload shifting.
    }}
    \label{fig4}
\end{figure}

The effectiveness of temporal workload shifting for peak power demand shaving under different delay rates is illustrated in Fig. \ref{fig4}, and the flexibility window is set to 15:00--15:25. The original data center power profile exhibits substantial temporal variability. As $\gamma$ increases, a wider temporal shifting range become available for flexible workloads, enabling greater reductions in power consumption. The enlarged view around the peak period further illustrates that the optimized profiles remain below the original demand during the flexibility window. These results demonstrate that temporal workload flexibility can effectively mitigate short term peak demand without directly curtailing workload execution.

Two cases are considered to evaluate the impact of delay rate on workload flexibility. Case 1 considers only GPU power demand, whereas Case 2 incorporates the power consumption of CPU, GPU and memory resources. As shown in Table \ref{tab:gamma_flexibility}, Case 2 consistently achieves higher flexibility across all delay rates. In particular, at \(\gamma=1.0\), the available flexibility increases from \(21.71\) kW in Case 1 to \(32.47\) kW in Case 2, representing a \(49.6\%\) increase. Therefore, Case 2 provides a more comprehensive representation of workload power demand.

Table \ref{tab:gamma_flexibility} also shows that increasing the delay rate \(\gamma\) generally enhances the available workload flexibility, with \(F\) increasing from 10.76 kW at \(\gamma=0\) to 32.47 kW at \(\gamma=1.0\) in Case 2. This corresponds to an increase in flexibility from about 2\% to 7\% of the approximately 540 kW total demand. This improvement is accompanied by a larger number of shifted workloads $n_w$ and, at higher $\gamma$, a longer average shifting time $t_\mathrm{shift}$, indicating that greater scheduling freedom enables more effective load shaving but requires more extensive temporal adjustment of workloads.
\begin{table}[t]
\centering
\caption{Impact of Delay Rate on Data Center Workload Flexibility}
\label{tab:gamma_flexibility}
\begin{tabular}{c ccc ccc}
\hline
\multirow{2}{*}{$\gamma$}
& \multicolumn{3}{c}{Case 1\cite{caprara2026data}}
& \multicolumn{3}{c}{Case 2} \\
\cline{2-4}\cline{5-7}
& $F$ (kW) & $n_w$ & $t_{\mathrm{shift}}$ (min)
& $F$ (kW) & $n_w$ & $t_{\mathrm{shift}}$ (min) \\
\hline
0   & 8.690 & 33 & 97.42 & 10.76 & 37 & 90.27  \\
0.2 & 14.35 & 48 & 57.08 & 19.01 & 49 & 57.14  \\
0.4 & 17.35 & 64 & 64.14 & 25.86 & 70 & 61.29  \\
0.6 & 19.96 & 64 & 88.05 & 28.95 & 71 & 82.54  \\
0.8 & 21.16 & 68 & 119.0 & 31.17 & 77 & 115.7  \\
1.0 & 21.71 & 69 & 147.1 & 32.47 & 78 & 147.4  \\
\hline
\end{tabular}
\end{table}

\begin{figure}[t]
    \centering
    \includegraphics[width=3.5in]{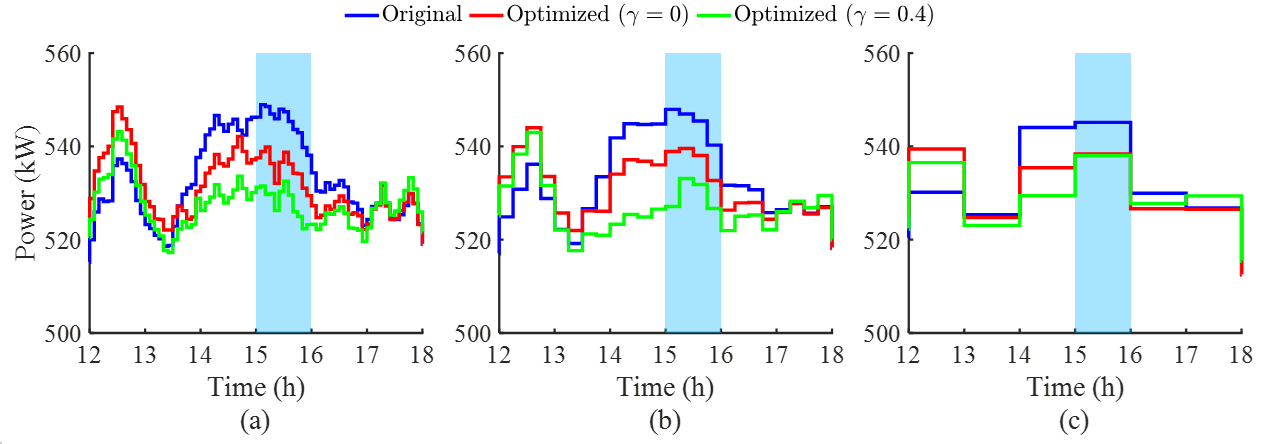}
    \caption{{
    \raggedright
    The power load profiles under different temporal resolutions. (a) 5 min, (b) 15 min, (c) 60 min
    }}
    \label{diff reso}
\end{figure}

We further test the workload flexibility under different time resolutions, as shown in Fig. \ref{diff reso}, with the flexibility window set to 15:00--16:00. As the temporal resolution becomes coarser, the maximum achievable flexibility decreases because fewer workloads can be effectively shifted. This result indicates that finer temporal resolutions preserve more workload scheduling opportunities and enable greater demand-side flexibility.

\subsection{Sustained Power Demand Reduction}
% Sustained power demand reduction is further considered to assess whether the available workload flexibility can be maintained over different time durations. This is important for grid services that require continuous demand reduction rather than a short-duration response, and it provides a more comprehensive measure of the duration-dependent flexibility of the data center.
The maximum duration under different power reduction requirements and delay rates are shown in Table \ref{tab:sustained_duration}. For a fixed delay rate, increasing the required power reduction shortens the duration for which that reduction can be continuously maintained. For example, under \(\gamma=0.4\), the maximum duration \(D_{\max}\) decreases from 340 min at \(F^{\mathrm{req}}=10\) kW to 195 min at 15 kW and further to 120 min at 20 kW. A similar trend is observed for \(\gamma=0\), however, \(F^{\mathrm{req}}=20\) exceeds the maximum workload flexibility of data center. These results highlight an inherent magnitude–duration tradeoff in workload-based data center flexibility.

\begin{table}[t]
\centering
\caption{Maximum Duration Under Different Power Reduction Requirements}
\label{tab:sustained_duration}
\begin{tabular}{c c c c c c}
\hline
$F^{\mathrm{req}}$ (kW) & $\gamma$ & Sustained Interval 
& $D_{\max}$ (min) & $n_w$ & $t_{\mathrm{shift}}$ (min) \\
\hline

\multirow{2}{*}{10}
& 0   & 20:10--22:40 & 150  & 41  & 74.76    \\
& 0.4 & 09:55--15:35 & 340 & 121 & 90.74 \\
\hline

\multirow{2}{*}{15}
& 0   & 21:45--22:35   & 50  & 19  & 109.7    \\
& 0.4 & 12:25--15:40 & 195 & 224 & 51.09 \\
\hline

\multirow{2}{*}{20}
& 0   & --           & --  & --  & --    \\
& 0.4 & 13:30--15:30 & 120 & 68  & 81.40 \\
\hline

\end{tabular}
\end{table}

% \subsection{Computational Time}
% The mean solution times for the 5-min, 15-min, 1-h, and 2-h flexibility windows are approximately 2.06 s, 2.00 s, 2.03 s, and 2.18 s, respectively. The solution time remains relatively stable across different flexibility durations. This indicates that the proposed optimization framework maintains consistent computational efficiency over different sustained flexibility requirements.
\section{Conclusions}
This paper proposes a workload-level framework for assessing the grid-compatible demand flexibility of AI data centers through batch workload shifting. An averaging-based data processing method is developed to unify fine-resolution CPU, GPU, and memory traces with power system scheduling time intervals, and the workload model is coupled with server resource and power constraints. Numerical results show that greater workload delay tolerance increases short-term peak-shaving capability, while finer temporal resolutions preserve more scheduling flexibility. The sustained reduction analysis further reveals a clear tradeoff between the required power reduction and its maximum achievable duration. Overall, the proposed framework provides a practical means to quantify both the magnitude and duration of AI data center demand flexibility for power system operation.

\bibliographystyle{IEEEtran}
\bibliography{main}

\end{document}